\documentclass[11pt]{article}

\usepackage{a4wide}
\usepackage[pdftex,usenames,dvipsnames]{color}
\usepackage{graphics}
\usepackage{amsfonts}
\usepackage{amssymb}

\newcommand{\nit}{\noindent}

\newcommand{\np}{\newpage}
\newcommand{\dsp}{\displaystyle}
\newcommand{\vs}[1]{\vspace{#1 ex}}
\newcommand{\hs}[1]{\hspace{#1 em}}
\newcommand{\bfr}{\begin{flushright}}
\newcommand{\efr}{\end{flushright}}
\newcommand{\bc}{\begin{center}}
\newcommand{\ec}{\end{center}}
\newcommand{\ben}{\begin{enumerate}}
\newcommand{\een}{\end{enumerate}}

\newcommand{\be}{\begin{equation}}
\newcommand{\ee}{\end{equation}}
\newcommand{\ba}{\begin{array}}
\newcommand{\ea}{\end{array}}
\newcommand{\ct}{\cite}
\newcommand{\bit}{\bibitem}

\newcommand{\del}{\delta}
\newcommand{\eps}{\epsilon}
\newcommand{\ve}{\varepsilon}

\newcommand{\lb}{\lambda}

\newcommand{\vf}{\varphi}
\newcommand{\og}{\omega}

\newcommand{\Del}{\Delta}

\newcommand{\Sg}{\Sigma}

\newcommand{\bfx}{\bold{x}}

\newcommand{\lh}{\left(}
\newcommand{\rh}{\right)}
\newcommand{\ld}{\left.}
\newcommand{\rd}{\right.}

\newcommand{\nb}{\nabla}
\newcommand{\bfnb}{\mbox{\boldmath{$\nb$}}}

\newcommand{\ctgh}{\mbox{\,cotanh\,}}

\newcommand{\der}{\partial}

\begin{document}

\pagestyle{empty} 
\bfr
NIKHEF-2013-001
\efr

\begin{center} 
{\Large{\bf On single scalar field cosmology}} \\
\vs{7} 

{\large J.W.\ van Holten}\\ 
\vs{5}

{\large NIKHEF}
\vs{2} 

{\large Amsterdam NL} \\ 
\vs{7}

January 4, 2013
\end{center} 
\vs{5}

\nit
{\small
{\bf Abstract} \\
Observations suggest, that there may be periods in the history of the universe, including the 
present one, in which its evolution is driven by scalar fields. This paper is concerned with 
the solution of the evolution equations for a spatially flat universe driven by a single scalar field. 
Some general theorems relevant to the cosmology of these models are presented, and several 
approaches to solve the equations are discussed. For some potentials special exact solutions can 
be found, and for the case of exponential potentials the complete solution is rederived in a new
parametrization. For the general case solutions are constructed in terms of a power series 
expansion in the field. The issue of double-valuedness of such a series expansion in case of 
oscillating fields with turning points is addressed and resolved. 
}
 
\np
~\hfill

\np

\pagestyle{plain}
\pagenumbering{arabic} 

\section{Cosmic scalar fields \label{s1}} 

It is widely accepted that scalar fields can drive the cosmic expansion. More in particular, scalar fields 
can possibly account both for an early period of inflation to explain the large-scale homogeneity and 
isotropy of the universe \ct{starobinsky1980,guth1981}, and for the observed accelerated expansion 
of the universe in more recent times \ct{riess_etal1998, perlmutter_etal1999}. One of the most widely studied 
scenarios for inflation is the minimal chaotic inflation model \ct{linde1983}, in which a scalar field moving 
in a potential creates a dynamical form of dark energy that makes the universe expand. In a similar way 
cosmic scalar fields can be used to model dark energy \ct{wetterich1988, zlatev_etal1999} which drives 
the accelerated expansion of the universe deduced from supernova observations. In view of such potential 
applications the dynamics of scalar fields in a cosmological context is a physically relevant subject of 
investigation \ct{turner1983,vantent2002}. 

In this paper we investigate simple cosmological models in which a single dynamical scalar is minimally 
coupled to gravity, and the gravitational field is taken to be of the Friedmann-Lemaitre-Robertson-Walker 
type. More in particular, motivated by the observations of the cosmic microwave background we take the 
spatial part of the metric to be flat. Thus the line element describing the proper-time for a comoving particle 
in the universe is\footnote{In most of this paper we employ Planck units such that $c = \hbar = 8 \pi G = 1$.}
\be
- d\tau^2 = g_{\mu\nu}\, dx^{\mu} dx^{\nu} = - N^2(t)\, dt^2 + a^2(t)\, d\bfx^2.  
\label{1.1}
\ee
Here $a(t)$ is the scale factor, whilst $N(t)$ is the lapse function in the ADM formulation of General Relativity 
\ct{adm}, allowing one to keep local time reparametrizations as an invariance in the description of space-time  
geometry. Usually one chooses $t$ to be cosmic time such that $N(t) = 1$. This will also be our prefered 
choice. However, we find it useful to keep $N(t)$ free in the derivation of the relevant field equations for 
reasons to become clear soon. 

The action for a real scalar field $\vf$ minimally coupled to gravity in General Relativity is 
\be
S = \int d^4x\, \sqrt{-g} \lh - \frac{1}{2}\, R - \frac{1}{2}\, g^{\mu\nu}\, \der_{\mu} \vf\, \der_{\nu} \vf - V[\vf] \rh,
\label{1.2}
\ee
where $R$ is the Riemann scalar and $V[\vf]$ is the scalar potential. Taking the metric of the form (\ref{1.1}) 
and the scalar field to be spatially constant in this frame: $\bfnb \vf = 0$,  $\vf(x^{\mu}) = \vf(t)$, the effective 
action per unit of co-ordinate volume takes the form 
\be
\Sg = \int dt \lh - \frac{3}{N}\, a \dot{a}^2 + \frac{a^3}{2N}\, \dot{\vf}^2 - a^3 N V[\vf] \rh.
\label{1.3}
\ee
Variation of this action provides the relevant equations of motion for the evolution of this isotropic and 
homogeneous model universe. First, the dynamical equation for the scale factor is 
\be
\frac{1}{3aN} \frac{\del \Sg}{\del a} = \frac{2}{aN} \frac{d}{dt} \lh \frac{1}{N} \frac{da}{dt} \rh +
 \lh \frac{1}{aN} \frac{da}{dt} \rh^2 + \frac{1}{2} \lh \frac{1}{N} \frac{d\vf}{dt} \rh^2 - V[\vf] = 0. 
\label{1.4.1}
\ee
Next, the dynamical equation for the scalar field is
\be
- \frac{1}{a^3 N}\, \frac{\del \Sg}{\del \vf} = \frac{1}{N} \frac{d}{dt} \lh \frac{1}{N} \frac{d\vf}{dt} \rh 
 + \lh \frac{3}{aN} \frac{da}{dt} \rh \frac{1}{N} \frac{d\vf}{dt} + V^{\prime}[\vf] = 0,
\label{1.4.2}
\ee
where the prime denotes a derivative w.r.t.\  the field $\vf$. Finally, there is a constraint imposed by
variation of the arbitrary lapse function $N$:
\be
\frac{1}{a^3} \frac{\del \Sg}{\del N} = 3 \lh \frac{1}{aN} \frac{da}{dt} \rh^2 - \frac{1}{2} \lh \frac{1}{N} \frac{d\vf}{dt} \rh^2
 - V[\vf] = 0.
\label{1.4.3}
\ee
In this form, both the effective action and the equations of motion are manifestly invariant under time
reparametrizations $t \rightarrow t^{\prime}$, with $N(t)$ transforming as 
\be
N^{\prime} (t^{\prime}) dt^{\prime} = N(t) dt,
\label{1.5}
\ee
whilst $a(t)$ and $\vf(t)$ behave as scalars: 
\be
a^{\prime}(t^{\prime}) = a(t), \hs{2} \vf^{\prime}(t^{\prime}) = \vf(t).
\label{1.6}
\ee
This reparametrization invariance allows us to choose a gauge $N = 1$, such that the equations of motion
become
\be
\ba{l}
\dsp{ 2 \dot{H} + 3 H^2 + \frac{1}{2}\, \dot{\vf}^2 - V = 0, }\\ 
 \\
\dsp{ \ddot{\vf} + 3 H \dot{\vf} + V^{\prime} = 0, }\\
 \\
\dsp{ - 3 H^2 + \frac{1}{2}\, \dot{\vf}^2 + V = 0, }
\ea
\label{1.7}
\ee
where $H = \dot{a}/a$ is the usual Hubble parameter. 
It is well-known that these equations are redundant to the extent that for $H \neq 0$ the last two equations 
imply the first one; indeed, for $H \neq 0$ differentiation of the last equation leads --upon use of the middle 
one-- to
\be
6 H \dot{H} + 3 H \dot{\vf}^2 = 0 \hs{1} \Rightarrow \hs{1}  2 \dot{H} = - \dot{\vf}^2.
\label{1.8}
\ee
Adding this to the last equation (\ref{1.7}) gives back the first one. However, when $H = 0$ the proof fails, 
and the first equation has to be considered as a separate condition. Actually, the last equation (\ref{1.7}) 
obtained from the variation of $N$ is just the reduced hamiltonian constraint of general relativity in 
FLRW space-times (also known as the Wheeler-DeWitt equation) which restricts the set of allowed 
solutions of the equations of motion to those for which the hamiltonian vanishes. 

In this context we observe, that the effective action (\ref{1.3}) can be written in a more familiar form by 
defining new dynamical variables $(X^0, X^1)$ and a time parameter $\tau(t)$ by
\be
X^0 = \sqrt{6} \ln a, \hs{2} X^1 = \vf, \hs{2} \frac{d\tau}{dt} = \frac{1}{a^3} = e^{- \sqrt{3/2}\, X^0}.
\label{1.9}
\ee
In terms of these variables the effective the effective action becomes \ct{kolb1995}
\be
\Sg = \int d\tau \lh - \frac{1}{2N} \lh \frac{dX^0}{d\tau} \rh^2 + \frac{1}{2N} \lh \frac{dX^1}{d\tau} \rh^2
 - N U[X^0,X^1] \rh,
\label{1.10}
\ee
where
\be
U[X^0,X^1] = e^{\sqrt{6}\, X^0} V[X^1] = a^6 V[\vf].
\label{1.11}
\ee
Thus the cosmological model (\ref{1.3}) is mathematically equivalent to that of a relativistic particle in Minkowski 
space moving in a time-dependent scalar potential $U$. The action (\ref{1.10}) is a convenient starting 
point for a canonical (and quantum) treatment of mini-superspace cosmology. In this formulation the 
hamiltonian constraint in the gauge $N = 1$ takes the simple form
\be
- \frac{1}{2} \lh \frac{dX^0}{d\tau} \rh^2 + \frac{1}{2} \lh \frac{dX^1}{d\tau} \rh^2 + U[X^0,X^1] = 0.
\label{1.12}
\ee

\section{Dynamics: general considerations \label{s2}}

Returning to the classical cosmology model described by eqs.\ (\ref{1.7}), we observe that the hamiltonian
constraint represents a first integral of motion for the system $(a, \vf)$,  but one which can not take arbitrary
values: being a first-class constraint the right-hand side of the third equation (\ref{1.7}) necessarily vanishes,
even though the other two equations would be consistent with any constant value $E$ such that
\be
a^3 \lh -3 H^2 + \frac{1}{2}\, \dot{\vf}^2 + V \rh = E.
\label{2.1}
\ee
The constraint $E = 0$ is the result of local time-reparametrization invariance imposed by the gauge 
variable $N$. It follows that one can not impose arbitrary initial conditions for the variables $(a(t), \vf(t))$ 
and their velocities $(\dot{a}(t), \dot{\vf}(t))$: any set of initial values is constrained by $E = 0$. Keeping 
this in mind, the complete classical dynamics can be derived from the hamiltonian constraint and the 
Klein-Gordon equation for $\vf$, i.e.\ the second eq.\ (\ref{1.7}). 

It is clear that the middle term of the KG equation linear in the velocity $\dot{\vf}$ represents 
the transfer of energy from the scalar field to the scale factor, or vice versa. Note however, that 
it does not imply the breaking of time reversal invariance: under time reversal $t \rightarrow -t$ 
both $\dot{\vf}$ and $H$ change sign, with the effect that the equation itself is time-reversal 
invariant. As a result the energy density $E/a^3$ remains constant (and vanishes), and there 
is no dissipation of energy for the combined system of scalar and gravitational degrees of freedom 
as such. To include dissipation by e.g.\ particle creation, the equations (\ref{1.7}) would have to be
modified \ct{turner1983}.

A related conclusion is, that consistent non-degenerate evolution of the system does not 
necessarily require the scalar potential $V$ to be bounded below: there is already a negative contribution
to the hamiltonian from gravity, witness the term $- 3 H^2$ in eq.\ (\ref{2.1}), but the constraint $E = 0$ 
serves to stabilize the system. 

As observed earlier the dynamics of gravity, as described by the first equation (\ref{1.7}), only follows
from the other two equations if $H \neq 0$; therefore in solving for $a$ and $\vf$ from the KG equation 
and the constraint we always have to consider the case $H = 0$ corresponding to flat space-time 
separately. The case of Minkowski space-time $H = \dot{H} = 0$ is quite straightforward: the 
equations (\ref{1.7}) reduce to 
\be
\frac{1}{2}\, \dot{\vf}^2 - V = \frac{1}{2}\, \dot{\vf}^2 + V = 0, \hs{2} \ddot{\vf} + V^{\prime} = 0.
\label{2.1.1}
\ee
It follows that the kinetic and potential energy have to vanish separately:
\be
\dot{\vf} = V = 0, 
\label{2.1.2}
\ee
and as a result also
\be 
\ddot{\vf} = V^{\prime} = 0.
\label{2.1.3}
\ee
This is possible only if the potential has a stationary point which is also a zero: $V = V^{\prime} = 0$. 
In most potentials this will not apply. Note, that of course the evolution of the universe can pass through 
a flat point, where $H = 0$ but $\dot{H} \neq 0$. At such a point 
\be
\dot{H} = V = - \frac{1}{2}\, \dot{\vf}^2 \leq 0,
\label{2.1.4}
\ee 
i.e.\ the universal expansion goes through a maximum there.

In all other cases $(H \neq 0)$, it suffices to consider the two equations
\[
\ddot{\vf} + 3 H \dot{\vf} + V^{\prime} = 0, \hs{2}
\frac{1}{2}\, \dot{\vf}^2 + V = 3 H^2. 
\]
Now assuming a solution to exist, this solution must at least for finite stretches of time define a 
one-to-one map from $t$ to $\vf$. During such a period one can write \ct{kolb1995,vholten2002}:
\be
H(t) = H[\vf(t)], \hs{2} \dot{H} = H^{\prime} \dot{\vf},
\label{2.2}
\ee
where the overdot denotes a derivative w.r.t.\ cosmic time $t$, and the prime a derivative w.r.t.\ the 
scalar field $\vf$. Eq.\ (\ref{1.8}) then implies
\be
\dot{\vf}^2 = - 2 H^{\prime} \dot{\vf},
\label{2.3}
\ee
hence either $\dot{\vf} = 0$ or 
\be
\dot{\vf} = - 2 H^{\prime}, \hs{2} \ddot{\vf} = 4 H^{\prime\prime} H^{\prime}.
\label{2.4}
\ee
After this is substituted back into the equations of motion we find \ct{copeland_etal1993,kolb1995,vholten2002}
\be
V = 3 H^2 - 2 H^{\prime\, 2}, \hs{2} V^{\prime} = 2 H^{\prime} \lh 3 H - 2 H^{\prime\prime} \rh.
\label{2.5}
\ee
It follows, that $\dot{\vf}$ and $H^{\prime}$ can never vanish when $V < 0$, and conversely 
a configuration in which $\dot{\vf} = H^{\prime} = 0$ can be reached only in regions where $V \geq 0$. 
It also follows, that stationary points of $V$ impose a particular constraint on $H[\vf]$:
\be
V^{\prime} = 0 \hs{1} \Rightarrow \hs{1} H^{\prime} = 0 \hs{1} \mbox{or} \hs{1} 2H^{\prime\prime} = 3H.
\label{2.5.1}
\ee
In contrast, a stationary point of $H$ will occur at a stationary point of $V$, unless $H^{\prime\prime}$ 
is singular there and $H^{\prime} H^{\prime\prime}$ is finite and non-zero. In the latter case, eqs.\ (\ref{2.4}) 
imply that $\dot{\vf} = 0$, $\ddot{\vf} \neq 0$; hence $\vf$ reaches an extremum and its trajectory in the 
$(t, \vf)$-plane exhibits a turning point there. 

\section{Explicit examples of scalar cosmology  \label{s2a}} 

There are several ways to construct solutions for the equations of scalar cosmology, depending on the
problem to be addressed. For example, one can search for models and initial conditions allowing for a 
specific type of solution, or one can try to find all solutions allowed by a given scalar potential and initial
conditions. In this section we discuss examples of the first type; in the later sections we construct more 
generic solutions for specific potentials. 

Various cosmological scenarios, such as purely matter or radiation dominated universes, are 
described by a simple power law:
\be
a(t) = \lh \frac{t}{\tau} \rh^n, 
\label{s2a.1}
\ee
where $\tau$ is some fixed reference time. Such types of behaviour can also be achieved in 
scalar field cosmology \ct{padmanabhan2002}; it follows from (\ref{s2a.1}) that
\be
H = \frac{\dot{a}}{a} = \frac{n}{t},
\label{s2a.2}
\ee
and the second equation (\ref{1.8}) becomes
\be
\dot{\vf}^2 = - 2 \dot{H} = \frac{2n}{t^2} \hs{1} \Rightarrow \hs{1} \dot{\vf} = \pm \frac{\sqrt{2n}}{t}.
\label{s2a.3}
\ee
Clearly a power-law solution (\ref{s2a.1}) is possible in the context of a regular scalar field model only 
for $n \geq 0$, i.e.\ non-contracting universes. Introducing a constant of integration $\tau$, the solution 
of eq.\ (\ref{s2a.3}) is \ct{padmanabhan2002,russo2004}
\be
\vf(t) = \vf(\tau) \pm \sqrt{2n} \ln \frac{t}{\tau}. 
\label{s2a.4}
\ee
Substitution of these results into the third eq.\ (\ref{1.7}) now leads to
\be
V = 3 H^2 - \frac{1}{2}\, \dot{\vf}^2 = \frac{n(3n - 1)}{t^2} = V_0\, e^{\mp \sqrt{\frac{2}{n}}\, \vf},
\label{s2a.5}
\ee
with $V_0$ fixed by the requirement
\be
V_0\, e^{\mp \sqrt{\frac{2}{n}}\, \vf(\tau)} = \frac{n \lh 3n - 1 \rh}{\tau^2 }.
\label{s2a.6}
\ee
Observe, that for $0 < n < 1/3$ the potential is negative definite, whereas for $n > 1/3$ it is positive 
definite. We conclude, that both positive and negative exponential potentials can allow for power-law 
solutions of the type (\ref{s2a.1}), but only in specific domains of non-negative powers $n$. We will 
show later, that flat contracting universes can arise in scalar cosmology for other types of potentials, 
although not with a simple power-law (\ref{s2a.1}) for the scale factor. A different question to be 
addressed later is, what other solutions exist for exponential potentials. 

It goes without saying, that a constant Hubble parameter $H_0$ results from constant $\vf = \vf_0$ 
at an extremum of the potential:
\be
\dot{\vf} = 0, \hs{2} V'(\vf_0)= 0. 
\label{s2a.8}
\ee 
In scalar cosmology this necessarily represents a positive cosmological constant $V = 3 H_0^2$. 
Of course, a solution (\ref{s2a.8}), as well as the Minkowski limit $H_0 = 0$, can also arise as the 
final stationary state in a scenario in which the scalar field evolves dynamically from a higher value 
to end up at an extremum of the potential. 
\vs{1}

\nit
Instead of specifying a certain evolution of the scale factor, one can also start from a specification 
of the time-dependence of the scalar field. As an example, we construct a solution with an oscillating 
scalar field
\be
\vf(t) = \vf_0 \cos \og t.
\label{s2a.10}
\ee
The rate of change of the field is
\be
\dot{\vf} = - \og \vf_0 \sin \og t = - \og \sqrt{ \vf^2_0 - \vf^2}.
\label{s2a.11}
\ee
The first eq.\ (\ref{2.4}) then becomes
\be
H' = \frac{\og}{2} \sqrt{\vf^2_0 - \vf^2},
\label{s2a.12}
\ee
with the solution
\be
H = H_0 - \frac{\og \vf_0^2}{4}\, \arccos \frac{\vf}{\vf_0} + \frac{1}{4}\, \og \vf \sqrt{\vf_0^2 - \vf^2}. 
\label{s2a.13}
\ee
Here $H_0$ is the initial value of $H$ when $\vf = \vf_0$; indeed, substitution of (\ref{s2a.10}) gives 
the time dependence of $H$ as
\be
H(t) = H_0 - \frac{1}{4}\, \vf_0^2\, \og^2 t + \frac{1}{8}\, \vf_0^2\, \og \sin 2 \og t.
\label{s2a.14}
\ee
The corresponding solution for the scale factor reads
\be
a(t) = a_0 e^{H_0 t - \frac{1}{8}\, \vf_0^2 \og^2 t^2 + \frac{1}{16} \vf_0^2 \lh 1 - \cos 2\og t \rh}.
\label{s2a.15}
\ee
This represents a universe growing from very small size at large negative times, to a finite size around 
$t = 4H_0/(\og \vf_0)^2$, when it shows some oscillating behaviour, to contract again to arbitrarily small 
size for very large positive times. Finally, we can compute the scalar potential from which such behaviour 
follows:
\be
\ba{lll}
V & = & 3H^2 - 2 H^{\prime\, 2} \\
 & & \\
 & = & \dsp{ 3 \lh H_0 - \frac{\og \vf_0^2}{4}\, \arccos \frac{\vf}{\vf_0} + 
  \frac{1}{4}\, \og \vf \sqrt{\vf_0^2 - \vf^2} \rh^2 - \frac{\og^2}{2} \lh \vf_0^2 - \vf^2 \rh. }
\ea
\label{s2a.16}
\ee
It is rather remarkable that such a complicated scalar potential can give rise to a simple periodic 
solution for the field $\vf(t)$, and allows a complete solution for the scale factor. More importantly, 
we observe that this solution of the scalar cosmology equations clearly shows essentially reversible 
behaviour, and illustrates explicitly that the back reaction of the space-time curvature on the scalar 
field in the Klein-Gordon equation can not be interpreted off-hand as a dissipative friction term.

\section{More on exact solutions \label{s4}}

So far we have constructed potentials starting from a prescribed time-dependence of the scale factor 
$a(t)$ or the scalar field $\vf(t)$. There is yet another way of finding solutions for scalar cosmology 
models, allowing to construct particular solutions as well as generic ones. This method starts not from 
a prescribed time behaviour of the cosmological degrees of freedom, but from postulating the relation 
$H[\vf]$ in eq.\ (\ref{2.2}). In this section we only consider particular cases with simple analytic solutions. 
The construction of generic solutions is discussed later. 

We start with the simplest non-trivial example, in which $H$ is linear in the field $\vf$:
\be
H = h_0 + h_1 \vf, \hs{2} H^{\prime} = h_1.
\label{s4a.1}
\ee
It follows directly, that
\be
V = 3 \lh h_0 + h_1 \vf \rh^2 -  2 h_1^2.
\label{s4a.2}
\ee
To diagonalize the mass term, we need to make the shift
\be
\psi = \vf + \frac{h_0}{h_1} \hs{1} \Rightarrow \hs{1} V = - 2 h_1^2 + 3 h_1^2 \psi^2, \hs{2} H = h_1 \psi.
\label{s4a.3}
\ee
Now $h_1$ is directly porportional to the mass:
\be
m^2 = 6 h_1^2 \hs{1} \Rightarrow \hs{1} V = - \frac{1}{3}\, m^2 + \frac{1}{2}\, m^2 \psi^2, 
\hs{2} H = \frac{m}{\sqrt{6}}\, \psi.
\label{s4a.4}
\ee
The corresponding solutions for the field and scale factor are easily found:
\be
\ba{l}
\dsp{ \dot{\psi} = -2 H_{\psi} \hs{1} \Rightarrow \hs{1} \psi(t) = - \frac{2m}{\sqrt{6}} \lh t - t_0 \rh, }\\
 \\
\dsp{ H = - \frac{m^2}{3} \lh t - t_0 \rh \hs{1} \Rightarrow \hs{1} 
 a(t) = a_0\, e^{- \frac{m^2}{3} \lh t - t_0 \rh^2}. }
\ea
\label{s4a.5}
\ee
This is not the only solution for the quadratic potential $V$ in (\ref{s4a.4}), in fact it is a very special one:
the only solution which exists for all times; but it pays to consider it a bit more in detail. A first observation 
is, that the rate of change of the scalar field $\dot{\psi}$ is constant over the whole time domain 
$(-\infty, + \infty)$. Thus there is no dissipation of kinetic energy, in spite of the fact that $H$ does not 
vanish except at $t = t_0$. Secondly we observe, that the Hubble parameter $H(t)$ is negative for times 
$t > t_0$; indeed this universe expands only during the epoch $t < t_0$ from arbitrarily small scales to a 
maximal size when $a(t) = a_0$ for $t = t_0$, and contracts again to vanishingly small size at large 
positive time $t > t_0$.

Actually, this behaviour for large times is generic for potentials with a negative minimum: $V_{min} < 0$. 
This follows from two general observations. First, eq.\ (\ref{1.8}):
\[
\dot{H} = - \frac{1}{2}\, \dot{\vf}^2 \leq 0,
\] 
implies that for regular kinetic terms of the scalar field the Hubble parameter is a non-increasing function
of time, and is constant only at stationary points of the field evolution: $\dot{\vf} = 0$. Second, as eq.\ 
(\ref{2.5}) shows, any point where the potential is negative must satisfy
\be
V < 0 \hs{1} \Rightarrow \hs{1} 2 H^{\prime\, 2} > 3 H^2 > 0,
\label{s4a.6}
\ee
and therefore $\dot{\vf} = - 2 H'$ can never vanish in the range where the potential is negative. As a 
result a negative minimum of the potential $V$, even if it is the absolute minimum, can never represent 
a stationary point of the dynamics; we conclude that the field never comes to rest at a negative value 
of the potential, and the Hubble parameter is a monotonically decreasing function of time as long as $V$
is negative. This conclusion is in agreement with the general result of ref.\ \ct{giambo_et-al}.

It is not difficult to construct more examples of exact solutions for polynomial potentials by a similar 
procedure. For example, taking
\be
H = h_0 + h_2 \vf^2, \hs{2} H' = 2 h_2 \vf,
\label{s4a.7}
\ee
we get a quartic potential 
\be
V = V_0 + \frac{m^2}{2}\, \vf^2 + \frac{\lb}{4}\, \vf^4, 
\label{s4a.8}
\ee 
with
\be
V_0 = 3 h_0^2, \hs{2} m^2 = 12 h_0 h_2 - 16 h_2^2, \hs{2} \lb = 12 h_2^2,
\label{s4a.9}
\ee
which implies
\[
V_0 = \frac{4\lb}{9} \lh 1 + \frac{3m^2}{4\lb} \rh^2.
\]
The corresponding particular time-dependent solutions of the Friedmann and Klein-Gordon equations are
\be
\vf(t) = \vf(0) e^{-\og t}, \hs{2} a(t) = a(0) e^{\sqrt{V_0/3}\, t + \frac{1}{8} \vf^2(0) \lh 1 - e^{-2\og t} \rh}, 
\label{s4a.10}
\ee
with  $\og^2 = 4 \lb /3$. The cosmology of this model was discussed in detail in ref.\ \ct{vholten2002}. 

\section{On a square root of the Friedmann equation \label{s5}}

In some cases it is possible to use the above procedure to construct the complete set of solutions in closed 
form. To simplify the discussion it is convenient to rescale the scalar field and define
\be
u(t) = \sqrt{\frac{3}{2}}\, \vf(t).
\label{s5a.1}
\ee
With this change of variable, the Friedmann equation (\ref{2.5}) and the first field equation (\ref{2.4}) become
\be
H^2 - H_u^2 = \frac{1}{3}\, V, \hs{2} \dot{u} = - 3 H_u,
\label{s5a.2}
\ee
employing the notation $H_u = dH/du$. Now suppose the potential is positive definite: $V > 0$, in some
domain of values of $u$. We can then introduce a function $K(u)$ defined by
\be
H = \pm \sqrt{\frac{V}{3}}\, \cosh K, \hs{2} H_u = \pm \sqrt{\frac{V}{3}} \lh K_u \sinh K 
 + \frac{V_u}{2V} \cosh K \rh.
\label{s5a.3}
\ee
Observe, that $H$ can have either sign, but once the sign is fixed it cannot change anymore during the 
subsequent evolution of the universe; in the following we focus on positive $H$ so as to describe an
expanding universe. 

After substitution into eq.\ (\ref{s5a.2}) and taking a square root, the function $K$ is then seen to satisfy
\be
K_u + \frac{V_u}{2V} \ctgh K = \pm 1.
\label{s5a.4}
\ee
Observe that the equation is odd in $K$, hence in this equation the two sign choices are related by 
$K \rightarrow -K$. Therefore it is sufficient to consider only the case with $+1$ on the right-hand side.
An example is provided by the exponential potential (\ref{s2a.5}) \ct{russo2004}-\ct{andrianov2011}:
\[
V = V_0\, e^{\lb u}, \hs{2} V_0 > 0,
\]
leading to a very simple equation for $K$:
\be
K_u + \frac{\lb}{2} \ctgh K = 1.
\label{s5a.5}
\ee
First consider the special case\footnote{The case $\lb = -2$ is obtained directly by the transformation
$u \rightarrow - u$ and $K \rightarrow -K$.}  $\lb = 2$, for which
\be
K_u = 1 - \ctgh K.
\label{s5a.5.1}
\ee
Observe, that $K_u$ cannot vanish anywhere, except in the limit $K \rightarrow \pm\, \infty$.
The general solution of these equations is given implicitly by
\be
\ba{l}
2K - e^{2K} = 4 (u-u_0),
\ea
\label{s5a.5.2}
\ee
for some constant of integration $u_0$. The Hubble parameter for an expanding universe is then 
determined by
\be
\ba{l}
\dsp{ H = \frac{1}{6} \sqrt{3V_0 e^{2u_0}}\lh e^{2K} + 1 \rh e^{-\frac{1}{4} \lh 2K + e^{2K} \rh}. }
\ea
\label{s5a.5.3}
\ee
The explicit time dependence can be obtained from the relation
\be
\ba{l}
\dsp{ \dot{K} = \sqrt{3V_0 e^{2u_0}}\, e^{- \frac{1}{4} \lh 2K + e^{2K} \rh}. }
\ea
\label{s5a.5.4}
\ee
The two equations (\ref{s5a.5.3}) and (\ref{s5a.5.4}) can be combined to write 
\be
\dsp{ 3H = \frac{1}{2} \lh e^{2K} + 1 \rh \dot{K} = \dot{K} - \dot{u}. }
\label{s5a.5.5}
\ee
It follows that there is a direct relation between $K$, $u$ and $a$:
\be 
a^3 e^{u - K} = \mbox{constant}. 
\label{s5a.5.6}
\ee
The constant defines a reference scale $a_0$ such that
\be
e^K = e^u \lh \frac{a}{a_0} \rh^{3}.
\label{s5a.5.7}
\ee
Using this result one can eliminate $K$ in terms of $a$ and $u$. In addition, it also allows us to 
calculate the total expansion factor in some period of evolution, as expressed by the number of $e$-folds: 
\[
N = \ln \frac{a_2}{a_1} = \frac{1}{3} \lh K_2 - K_1 - u_2 + u_1 \rh 
 = \frac{1}{12} \lh 2K_2 - 2K_1 + e^{2K_2} - e^{2K_1} \rh. 
\]
Similar results can be derived for $\lb = -2$.
\vs{1}

\nit
Having disposed of the cases for which $\lb^2 = 4$, we next turn to the generic case $\lb^2 \neq 4$. 
In terms of the initial condition $K_0 = K(u_0)$ such that
\be
e^{- 2K_0/\lb} = \lh 1 + \frac{\lb}{2} \rh e^{-K_0} - \lh 1 - \frac{\lb}{2} \rh e^{K_0},
\label{s5a.6}
\ee
the full solution is then given by
\be
K + \frac{\lb}{2} \ln \left| \lh 1 + \frac{\lb}{2} \rh e^{-K} - \lh 1 - \frac{\lb}{2} \rh e^K \right| 
 = \lh 1 - \frac{\lb^2}{4} \rh \lh u - u_0 \rh. 
\label{s5a.7}
\ee
Equivalently,
\be
e^{\lh \frac{\lb}{2} - \frac{2}{\lb} \rh \lh u - u_0 \rh} = 
 e^{2K/\lb} \left| \lh 1 + \frac{\lb}{2} \rh e^{-K} - \lh 1 - \frac{\lb}{2} \rh e^K \right|.
\label{s5a.7.1}
\ee
The corresponding expression for the Hubble parameter is 
\be 
\ba{lll}
H & = & \dsp{ \frac{1}{6} \sqrt{ 3V_0 e^{\lb u_0}} \lh e^K + e^{-K} \rh \left[ 
 e^{2K/\lb} \left| \lh 1 + \frac{\lb}{2} \rh e^{-K} - \lh 1 - \frac{\lb}{2} \rh e^K \right|
  \right]^{\frac{1}{\frac{4}{\lb^2} - 1}}. }
\ea
\label{s5a.8}
\ee
The pair of equations (\ref{s5a.7}) and (\ref{s5a.8}) represent the parametrized general solutions 
for $(u, H)$, with time eliminated in favor of the parameter $K$. A well-known special solution of 
this kind is one for which $K$ is constant:
\be
\ctgh K = \frac{2}{\lb}, \hs{2} K_u = 0,
\label{s5a.12}
\ee
which requires $|\lb| < 2$. The Hubble parameter is then given by 
\be
H = \pm \sqrt{\frac{V_0}{3}} \frac{e^{\lb u/2}}{\sqrt{1 - \lb^2/4}},
\label{s5a.13}
\ee
with opposite signs for an expanding or contracting universe. It is easy to check directly by substitution 
that this is a solution of the Friedmann equation. To construct the dynamics explicitly, observe that 
eq.\ (\ref{s5a.4}) implies that 
\be
H_u = \sqrt{\frac{V}{3}} \sinh K, \hs{2} \dot{u} = - 3 H_u = - \sqrt{3V} \sinh K.
\label{s5a.9}
\ee
For the special solution (\ref{s5a.13}) this leads to  the results
\be
\dot{u} = \mp\, \sqrt{ \frac{3 \lb^2 V_0}{4 - \lb^2} }\, e^{\lb u/2}, \hs{2}
H = \frac{4}{3\lb^2} \frac{1}{t - t_0}.
\label{s5a.13.1}
\ee
In the general case, by using eq.\ (\ref{s5a.7}) one finds
\be
\ba{lll}
\dsp{ \frac{2 \dot{K}}{\sqrt{3V_0 e^{\lb u_0}}} }& = & \dsp{ 
 \left[ \lh 1 + \frac{\lb}{2} \rh e^{-K} - \lh 1 - \frac{\lb}{2} \rh e^K \right] \times }\\
 & & \\
 & & \dsp{ \hs{1} \left[ e^{2K/\lb} \left| \lh 1 + \frac{\lb}{2} \rh e^{-K} - \lh 1 - \frac{\lb}{2} \rh e^K \right| 
 \right]^{\frac{1}{1 - \frac{4}{\lb^2}}}. } 
\ea
\label{s5a.11}
\ee
This result can be used again to derive a direct relation between $a$, $u$ and $K$. Indeed, 
eqs.\ (\ref{s5a.8}) en (\ref{s5a.11}) together imply
\be
3H = \frac{\lh e^{-K} + e^K \rh \dot{K}}{\lh 1 + \frac{\lb}{2} \rh e^{-K} - \lh 1 - \frac{\lb}{2} \rh e^K}
 =  \frac{2}{\lb} \lh \dot{K} - \dot{u} \rh.
\label{s5a.11.1}
\ee
For $\lb \rightarrow \pm 2$ this reproduces the results (\ref{s5a.5.5}). The relation (\ref{s5a.5.6}) now generalizes
to
\be
a^3 e^{\frac{2}{\lb} \lh u - K \rh} = \mbox{constant} \hs{1} \Rightarrow \hs{1} 
e^K = e^u \lh \frac{a}{a_0} \rh^{3\lb/2}.
\label{s5a.11.2}
\ee
It is straightforward to extend the construction above to potentials which are negative definite: 
$V \leq 0$. This allows us to parametrize $H$ as 
\be
H = \sqrt{\left| \frac{V}{3} \right|} \sinh Q, \hs{2} 
H_u = \sqrt{\left| \frac{V}{3} \right|} \lh Q_u \cosh Q + \frac{V_u}{2V} \sinh Q \rh,
\label{s5a.14}
\ee
for some function $Q(u)$. In contrast to the previous case, eq.\ (\ref{s5a.3}), here there is no need 
of a sign choice, as it can be absorbed in the sign of $Q$. Moreover, $H$ can change sign during 
the evolution of the universe, in case $Q$ switches sign. We have noted before, that this is a 
fundamental difference between strictly non-negative potentials and potentials taking negative 
values in some domain of scalar field values. 

By taking a square root, the Friedmann equation becomes
\be
Q_u + \frac{V_u}{2V} \tanh Q = \pm 1,
\label{s5a.15}
\ee
where again the two sign choices are related by $Q \rightarrow -Q$ and we can restrict 
ourselves to the positive sign without loss of generality. 
Using the example of the exponential potential (\ref{s2a.5}), with negative amplitude:
\[
V = V_0 e^{\lb u}, \hs{2} V_0 < 0,
\]
eq.\ (\ref{s5a.15}) becomes
\be
Q_u + \frac{\lb}{2}\, \tanh Q = 1.
\label{s5a.16}
\ee
For $\lb^2 \neq 4$ the solution is 
\be
Q + \frac{\lb}{2} \ln \left| \lh 1 + \frac{\lb}{2} \rh e^{-Q} + \lh 1 - \frac{\lb}{2} \rh e^Q \right| 
 = \lh 1 - \frac{\lb^2}{4} \rh \lh u - u_0 \rh,
\label{s5a.17}
\ee
or equivalently
\be
e^{\lh \frac{2}{\lb} - \frac{\lb}{2} \rh \lh u - u_0 \rh} = e^{2 Q/\lb} \left| \lh 1 + \frac{\lb}{2} \rh e^{-Q} 
 + \lh 1 - \frac{\lb}{2} \rh e^Q \right|.
\label{s5a.18}
\ee
In this way we again construct a parametrized solution for the pair $(u, H)$.
For $|\lb| > 2$ there exists another special simple solution, with constant $Q$:
\be
\tanh Q = \frac{2}{\lb}, \hs{2} Q_u = 0.
\label{s5a.19}
\ee
The corresponding Hubble parameter is
\be
H = \pm \sqrt{\left| \frac{V_0}{3} \right|} \frac{e^{\lb u/2}}{\sqrt{ \frac{\lb^2}{4} - 1}}.
\label{s5a.20}
\ee
For the general solution
\be
H = \sqrt{\left| \frac{V}{3} \right|} \sinh Q, \hs{2} \dot{u} = - \sqrt{|3V|} \cosh Q,
\label{s5a.21}
\ee
and from (\ref{s5a.16}):
\be
\ba{lll}
\dot{Q} & = & \dsp{ \dot{u}\, Q_u = \sqrt{|3V|} \lh \frac{\lb}{2} \sinh Q - \cosh Q \rh = \frac{3\lb}{2} H + \dot{u}. }
\ea
\label{s5a.22}
\ee
This implies a relation between $Q$, $u$ and $a$ similar to (\ref{s5a.11.2}):
\be
e^Q = e^u \lh \frac{a}{a_0} \rh^{3\lb/2}.
\label{s5a.23}
\ee

\section{Series expansions: the regular case \label{s6}} 

For general potentials, even if one cannot produce exact solutions, one can always construct solutions 
for scalar cosmology based on a series expansion method. In this section we consider the regular case,
in which $\vf(t)$ is a single-valued function of time in some finite time domain. We have already observed
before, that in single-scalar cosmology $H(t)$ is a non-increasing function of time, and therefore it can
be represented by a well-behaved function $H[\vf(t)]$ in the time domain considered. Later we will also 
consider solutions in a time domain in which $\vf(t)$ has a turning point, and the equation for $H[\vf]$ 
becomes double-valued.

As in sect.\ \ref{s5}, it is convenient to work with a rescaled scalar field $u = \sqrt{3/2}\, \vf$, and
a Hubble parameter $H[u]$, satisfying the equations (\ref{s5a.2}). Let $u_0$ be a point in the 
regular domain; then we can develop $H$ in a power series
\be
\ba{lll}
H & = & \dsp{ \sum_{n \geq 0} h_n (u - u_0)^n = h_0 + h_1 (u-u_0) + h_2 (u-u_0)^2 + ... }\\
 & & \\
H_u & = & \dsp{ \sum_{n \geq 0} (n+1) h_{n+1} (u - u_0)^n = h_1 + h_2 (u-u_0) + ... }
\ea
\label{s6a.1}
\ee
Clearly, in this case 
\be
V_u = 6 H_u \lh H- H_{uu} \rh = 0 \hs{1} \Leftrightarrow \hs{1} H_u = 0.
\label{s6a.2}
\ee
i.e., $H$ can have a stationary point only at an extremum of the potential; in all other points $H[u]$ is
a strictly monotonically decreasing function of time, hence a monotonic function of $u$, decreasing or
increasing for positive or negative slope of $u(t)$, respectively:
\be
V_u \neq 0 \hs{1} \Rightarrow \hs{1} \left\{ 
   \ba{l}
   \dot{u} > 0 \hs{1} \Rightarrow \hs{1} H_u < 0, \\
     \\
   \dot{u} < 0 \hs{1} \Rightarrow \hs{1} H_u > 0.
   \ea \rd
\label{s6a.3}
\ee
For example, a quadratic potential 
\be
V = \ve + \frac{m^2}{3} u^2,
\label{s6a.4}
\ee
has a single minimum at $u = 0$. Hence $H$ can have a stationary point $H_u = 0$ only at $u = 0$, and
\be
H_u = 0 \hs{1} \Leftrightarrow \hs{1} H^2 = \frac{\ve}{3}.
\label{s6a.5}
\ee
This condition can only be fulfilled if $V(0) = \ve \geq 0$, and the stationary solution is a Minkowski space for
\be
u = 0, \hs{2} \ve = H = 0,
\label{s6a.6}
\ee
whilst it becomes a de Sitter space if
\be 
u = 0, \hs{2} \ve = 3 H^2 > 0.
\label{s6a.7}
\ee
There is no stationary solution for $\ve < 0$. Note, that all other solutions for any $\ve$ are non-stationary, 
with $\dot{u} = - 3 H_u \neq 0$.

Returning to the series expansion (\ref{s6a.1}), and a similar expansion for the potential:
\be 
V = \sum_{n \geq 0} v_n (u - u_0)^n = v_0 + v_1 (u - u_0) + v_2 (u-u_0)^2 + ...,
\label{s6a.8}
\ee
eqs.\ (\ref{s5a.2}) give rise to the infinite set of relations 
\be
\sum_{k = 0}^n \left[ h_k h_{n-k} - (k+1) (n-k+1) h_{k+1} h_{n-k+1} \right] = \frac{v_n}{3}, 
\label{s6a.9}
\ee
for all non-negative integers $n = 0, 1, 2, ...$, and to
\be
\dot{u} = - 3 \sum_{n \geq 0} (n+1) h_{n+1} (u - u_0)^n.
\label{s6a.10}
\ee
The first few relations (\ref{s6a.9}) in explicit form are
\be 
\ba{ll}
n = 0: & v_0 = 3 h_0^2 - 3 h_1^2, \\
 & \\
n = 1: & v_1 = 6 h_1 \lh h_0 - 2 h_2 \rh, \\
 & \\
n = 2: & v_2 = 3 h_1 \lh h_1 - 6 h_3 \rh + 6 h_2 \lh h_0 - 2h_2 \rh, \\
 & \\
n = 3: & v _3 = 6h_1 \lh h_2 - 4 h_4 \rh + 6 h_3 \lh h_0 - 6 h_2 \rh.
\ea
\label{s6a.11}
\ee
It follows that either $h_1 = H_u(u_0) = 0$, which can happen only  if $v_1 = V_u(u_0) = 0$
and $\dot{u}(u_0) = 0$, or $h_1 \neq  0$ and
\be
\ba{l}
\dsp{ h_0^2 = h_1^2 + \frac{v_0}{3}, \hs{2} 
h_2 = \frac{h_0}{2} \lh 1 - \frac{v_1}{6h_0h_1} \rh, }\\
 \\
\dsp{ h_3 = \frac{h_1}{6} \lh 1 + \frac{h_0 v_1}{6h_1^3} - \frac{v_1^2}{36 h_1^4} - \frac{v_2}{3h_1^2} \rh, }\\
 \\
\dsp{ h_ 4 = \frac{h_0}{24} - \frac{v_3}{24 h_1} - \frac{h_0}{72 h_1^2} 
 \lh 1 - \frac{v_1}{4h_0h_1} \rh \lh \frac{h_0v_1}{h_1} - \frac{v_1^2}{6h_1^2} - 2 v_2 \rh. }
\ea
\label{s6a.12}
\ee
Using these last results, the equation for the scalar field becomes
\be 
- \frac{\dot{u}}{3h_1} = 1 + \lh \frac{h_0}{h_1} - \frac{v_1}{6h_1^2} \rh (u - u_0) +
 \frac{1}{2} \lh 1 + \frac{h_0v_1}{6 h_1^3} - \frac{v_1^2}{36h_1^4} - \frac{v_2}{3h_1^2} \rh (u - u_0)^2 + ...
\label{s6.13}
\ee
Thus the complete solution is given in terms of two parameters $h_0 = H(u_0)$ and $h_1 = H_u(u_0)$, 
representing the initial conditions of the cosmology. 

A solution with $h_1 = 0$ exists only if $v_1 = 0$ and $u_0$ is an extremum of $V$; then eqs.\ (\ref{s6a.11}) 
reduce to
\be
\ba{ll}
v_0 = 3 h_0^2, & \dsp{ v_1 = 0, }\\
 & \\
v_2 = 6 h_0 h_2 - 12 h_2^2, & v_3 = 6 h_0 h_3 - 36 h_2 h_3, \\ 
 & \\ 
v_4 = 6 h_0 h_4 - 48 h_2 h_4 + 3 h_2^2 - 27 h_3^2, & ...
\ea
\label{s6a.14}
\ee
Therefore, if $v_2 \neq 0$ we have $H_{uu} = 2 h_2 \neq 0$, and the point $u_0$ is a point of inflection
of $H[u]$, where the field comes to rest (either momentarily or permanently). Obviously such solutions 
are very special, if they exist at all for some given potential. It requires a trajectory $H[u]$ to reach a 
local extremum of the potential at zero velocity. An example is provided by the special solution of the 
quartic potential in eq.\ (\ref{s4a.7}) and following, which comes to rest at the minimum $\vf_0 = 0$ of the 
potential (\ref{s4a.8}).
\vs{1}

\nit
From the results above one can also estimate the total expansion factor of the universe between two times 
$(t_1, t_2)$, as given by the number of $e$-folds. The central result is, that
\be 
N = \int_1^2 H dt = - \int_1^2 \frac{H}{3H_u} du = 
 - \frac{1}{3} \int_1^2  du\, \frac{h_0 + h_1 (u-u_0) + h_2 (u - u_0)^2 + ...}{h_1 + 2 h_2 (u - u_0) 
 + 3 h_3 (u - u_0)^2 + ...}
\label{s6a.15}
\ee
For the generic case $h_1 \neq 0$ the result is again a power series expansion
\be
N = \left. \sum_{k \geq 1} n_k (u - u_0)^k \right|_1^2 = 
 \left[ n_1 (u - u_0) + n_2 (u-u_0)^2 + n_3 (u-u_0)^3 + ... \right]_1^2,
\label{s6a.16}
\ee
with coefficients 
\be
\ba{l}
\dsp{ n_1 = - \frac{1}{3} \frac{h_0}{h_1}, \hs{2} n_2 = - \frac{1}{6} + \frac{1}{3} \frac{h_0 h_2}{h_1^2}, }\\
 \\
\dsp{ n_3 = \frac{1}{9} \frac{h_2}{h_1} + \frac{1}{3} \frac{h_0h_3}{h_1^2} - \frac{4}{9} \frac{h_0h_2^2}{h_1^3}, }\\
 \\
\dsp{ n_4 = \frac{1}{6} \frac{h_3}{h_1} + \frac{1}{3} \frac{h_0 h_4}{h_1^2} - \frac{1}{6} \frac{h_2^2}{h_1^2}
 - \frac{h_0h_2 h_3}{h_1^3} + \frac{2}{3} \frac{h_0 h_2^3}{h_1^4}. }
\ea
\label{s6a.17}
\ee
For the special case $h_1 = 0$ one gets in stead an expansion
\be
N = \left[ n_0 \ln (u - u_0) + n_1 (u - u_0) + n_2 (u-u_0)^2 + n_3 (u - u_0)^3 + ... \right]_1^2,
\label{s6a.17.1}
\ee
with
\be
\ba{l}
\dsp{ n_0 = - \frac{1}{6} \frac{h_0}{h_2}, \hs{2} n_1 = \frac{1}{4} \frac{h_0 h_3}{h_2^2}, }\\
 \\ 
\dsp{ n_2 = - \frac{1}{12} + \frac{1}{6} \frac{h_0 h_4}{h_2^2} - \frac{3}{16} \frac{h_0 h_3^2}{h_2^3}, }\\
 \\
\dsp{ n_3 = \frac{1}{36} \frac{h_3}{h_2} + \frac{5}{36} \frac{h_0 h_5}{h_2^2} - 
 \frac{1}{3} \frac{h_0 h_3 h_4}{h_2^3} + \frac{3}{16} \frac{h_0 h_3^2}{h_2^4}. }
\ea
\label{s6a.18}
\ee

\section{Applications \label{s7}}

In this section we apply the general results above to simple quadratic potentials (\ref{s6a.4}). 
The simplest models are those with $\ve = 0$, which have a Minkowski minimum $H = 0$
at $u = 0$. It is most convenient to expand around the minimum $u_0 = 0$. Then there is only 
one non-vanishing term in the potential
\be
v_2 = \frac{m^2}{3}, \hs{2} v_n= 0, \hs{1} n = 0, 1, 3, ...
\label{s7a.1}
\ee
As a result we get for the non-stationary solutions which all have $h_1 \neq 0$:
\be
\frac{h_0}{h_1} = \pm 1, \hs{1} \frac{h_2}{h_1} = \pm \frac{1}{2}, \hs{1} 
 \frac{h_3}{h_1} = \frac{1}{6} \lh 1  - \frac{m^2}{9 h_1^2} \rh, \hs{1} 
 \frac{h_4}{h_1} = \pm \frac{1}{24} \lh 1 + \frac{2m^2}{9h_1^2} \rh, \hs{1} ...
\label{s7a.2}
\ee
The power series expansion for $H[u]$ then takes the form
\be
\ba{lll}
H & = & \dsp{ \pm h_1 \left[ 1 \pm u + \frac{1}{2}\, u^2 \pm \frac{1}{6} \lh 1 - \frac{m^2}{9h_1^2} \rh u^3
  + \frac{1}{24} \lh 1 + \frac{2m^2}{9 h_1^2} \rh u^4 + ... \right] }\\
  & & \\
  & = & \dsp{ \pm h_1  \left[ e^{\pm u} + {\cal{O}} \lh m^2/h_1^2 \rh \right]. }
\ea
\label{s7a.3}
\ee
This result was to be expected, as in the limit $m^2 \rightarrow 0$ the potential vanishes and the solutions
of the Friedmann equation (\ref{s5a.2}) become pure exponentials. For the equation of motion of the scalar
field we get similarly
\be
- \frac{\dot{u}}{3h_1} = 1 \pm u + \frac{1}{2} \lh 1 - \frac{m^2}{9h_1^2} \rh u^2 + ...
 = e^{\pm u} + {\cal O}\lh m^2/h_1^2 \rh.
\label{s7a.4}
\ee

\bc
\scalebox{0.3}{\includegraphics{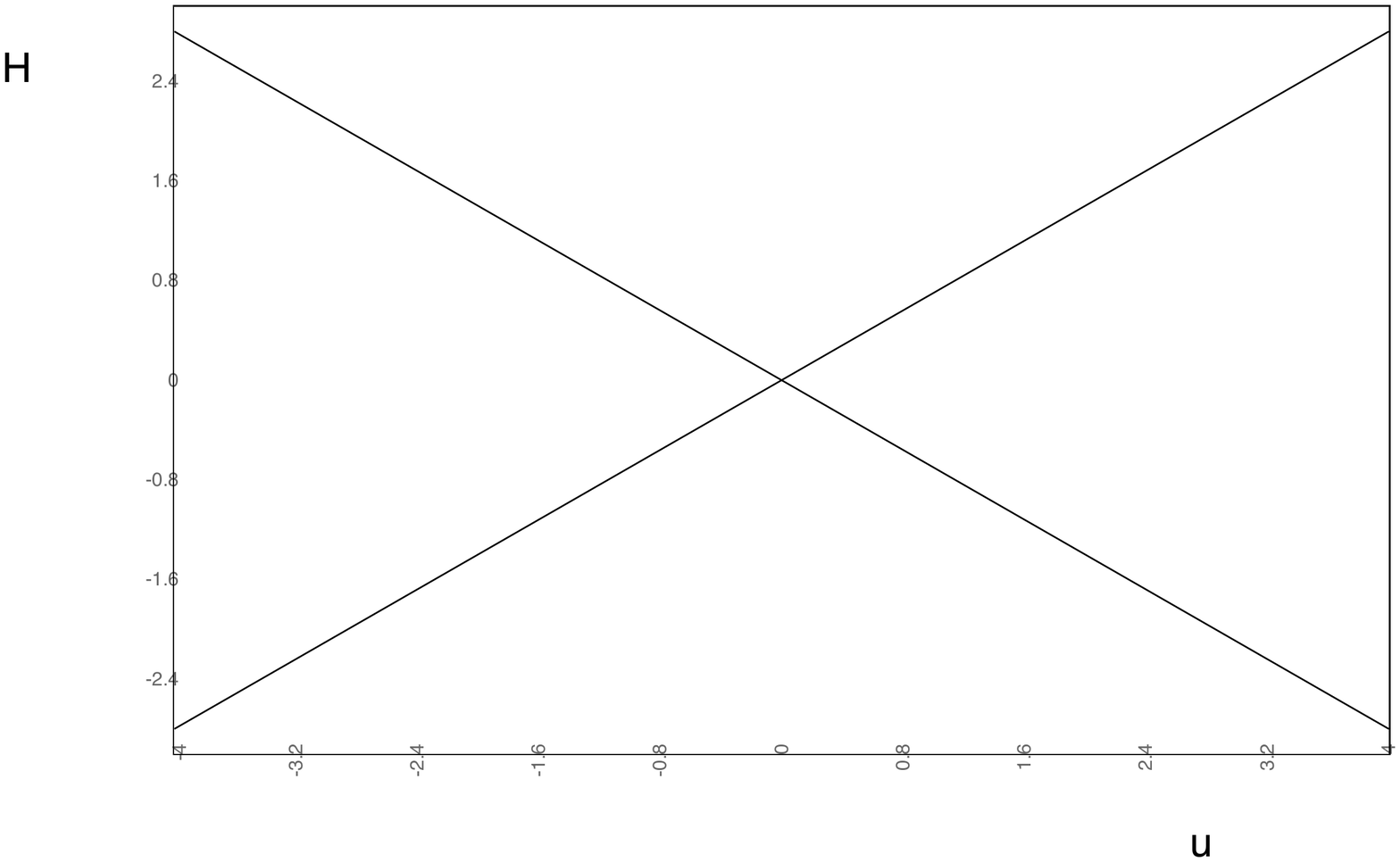}}

{\footnotesize{Fig.\ 1: Physical domain in the $(u,H)$-plane for $\ve = 0$ as determined by eq.\ (\ref{s7a.5}). }}
\ec

\nit
The domain of validity of this series expansion is restricted by the requirement that $H[u]$ is single valued. 
Now $u(t)$ can have a turning point only where $\dot{u} = - 3H_u = 0$. Therefore the locus of potential 
turning points is
\be 
H^2 = \frac{V}{3} = \lh \frac{mu}{3} \rh^2,
\label{s7a.5}
\ee
for the case at hand. Thus the $(u, H)$-plane is divided in four sectors by straight lines solving eq.\ 
(\ref{s7a.5}), and only the upper and lower quadrants in fig.\ 1 are allowed regions for the solutions (\ref{s7a.3}):
\be
- \left| \frac{mu}{3} \right|  \leq H[u] \leq \left| \frac{mu}{3}\right|.
\label{s7a.6}
\ee
There is no solution crossing from the upper to the lower quadrant. Indeed, at the point $u = H = 0$ where
the lines cross, $\dot{u} = H_u = 0$ and any solution passing through this point must have $h_0 = h_1 = 0$, 
hence $H_{uu}(0) = 2h_2 = 0$. It follows that no solution can pass from positive to negative $H$,
and the two sets of solutions are strictly separated. The only exceptional case is the Minkowski solution
represented by the point at the origin.
\vs{1}

\bc
\scalebox{0.3}{\includegraphics{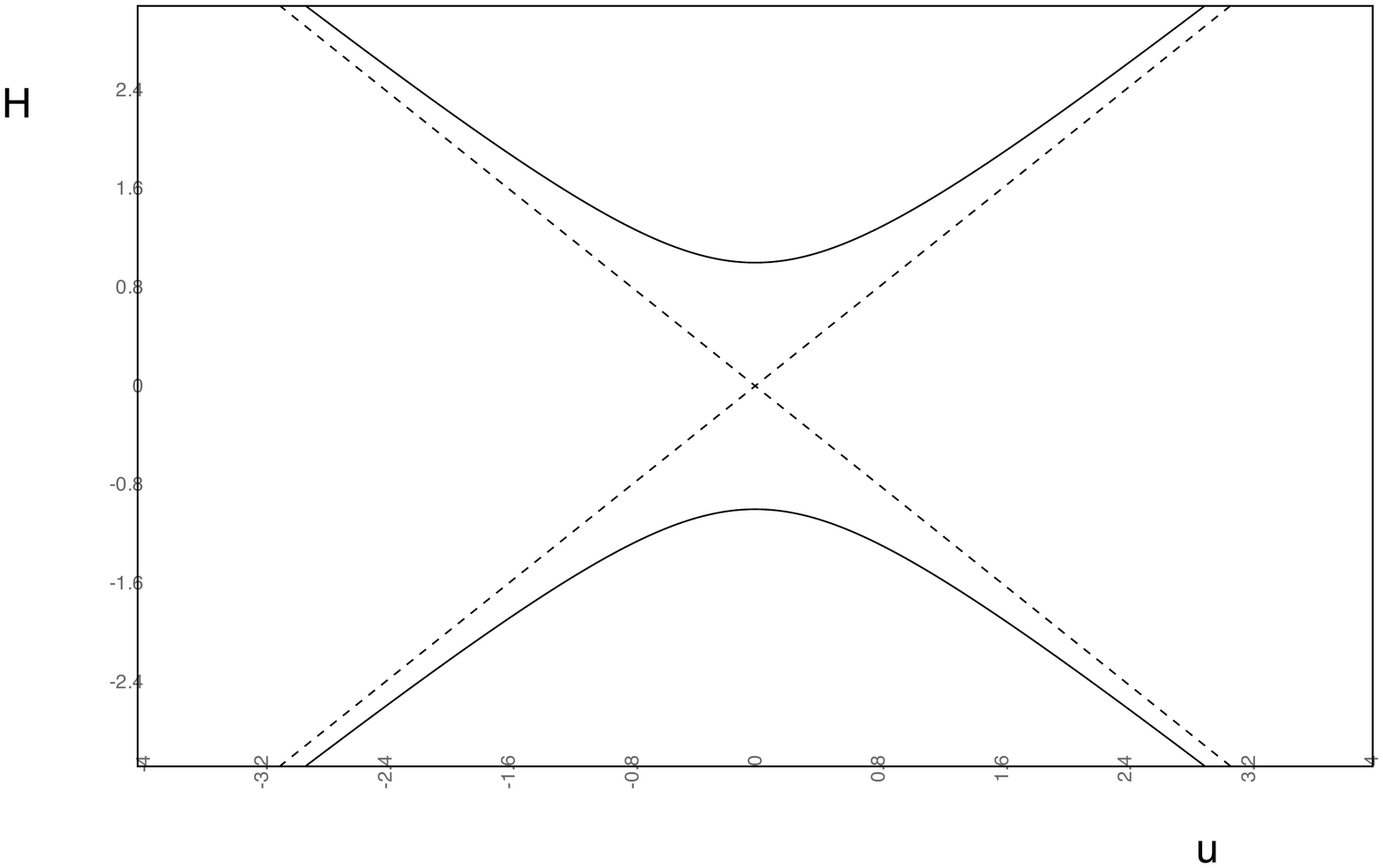}}

{\footnotesize{Fig.\ 2: Physical domain in the $(u,H)$-plane for $\ve > 0$ as determined by eq.\ (\ref{s7a.9})}}
\ec

\nit
Next we turn to the case $\ve > 0$, which has a stationary solution  $u = \dot{u} = 0$ representing de Sitter 
space, with a cosmological constant given by eq.\ (\ref{s6a.5}). The solution for $H$ is now slightly modified to
\be
\ba{ll}
\dsp{ \frac{h_0}{h_1} = \pm \sqrt{1 + \frac{\ve}{3h_1^2}}, }&
\dsp{ \frac{h_2}{h_1} = \frac{h_0}{2h_1} = \pm \frac{1}{2} \sqrt{1 + \frac{\ve}{3h_1^2}}, }\\
 & \\
\dsp{ \frac{h_3}{h_1} = \frac{1}{6} \lh 1 - \frac{m^2}{9h_1^2} \rh, }&
\dsp{ \frac{h_4}{h_1} = \pm \frac{1}{24} \sqrt{1 + \frac{\ve}{3h_1^2}} \lh 1 + \frac{2m^2}{9h_1^2} \rh, \hs{1} ... }
\ea
\label{s7a.7}
\ee
with the result
\be
\ba{rcl}
H & = & \dsp{ \pm h_1 \left[ \sqrt{1 + \frac{\ve}{3h_1^2}} \pm u + \frac{1}{2} \sqrt{1 + \frac{\ve}{3h_1^2}}\, u^2 
 \pm \frac{1}{6} \lh 1 - \frac{m^2}{9h_1^2} \rh u^3 \rd }\\
 \\
 & & \dsp{ \hs{2.5} \ld +\, \frac{1}{24} \sqrt{1 + \frac{\ve}{3h_1^2}}  \lh 1 + \frac{2m^2}{9h_1^2} \rh u^4 + ... \right], }\\
 \\
\dsp{ - \frac{\dot{u}}{3h_1} }& = & \dsp{ 1 \pm \sqrt{1 + \frac{\ve}{3h_1^2}}\, u 
 + \frac{1}{2} \lh 1 - \frac{m^2}{9h_1^2} \rh u^2 + ...}
\ea
\label{s7a.8}
\ee
In the $(u, H)$-plane the domain of validity of these approximations is restricted by the branches of the hyperbola
\be
H^2 - \lh \frac{m u}{3} \rh^2 = \frac{\ve}{3},
\label{s7a.9}
\ee
shown in fig.\ 2. On this hyperbola $\dot{u} = H_u = 0$. The two domains of allowed positive and negative 
$H$-values are separated by a gap of size $\Del H = 2 \sqrt{\ve/3}$. No cross-over is possible, and the 
solutions describe only permanently expanding or permanently contracting universes. 

\bc
\scalebox{0.3}{\includegraphics{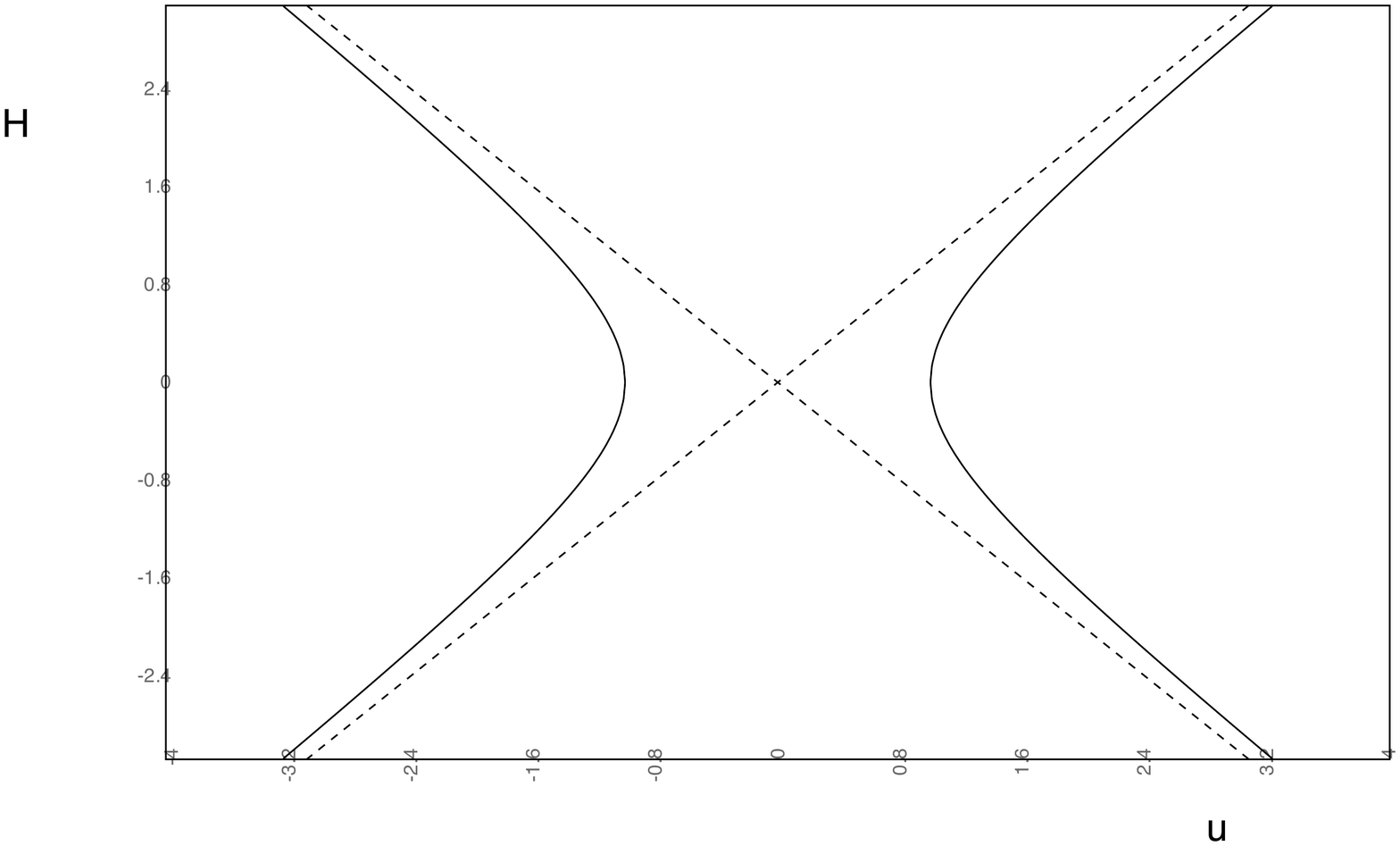}}

{\footnotesize{Fig.\ 3: Physical domain in the $(u,H)$-plane for $\ve < 0$ as determined by eq.\ (\ref{s7a.12}). }}
\ec

\nit
Finally, we consider the case $\ve < 0$. If there would be a solution $u = \dot{u} = 0$, this would give rise
to an anti-de-Sitter space. However, no such solution exists: in the domain $V < 0$ solutions of the 
Klein-Gordon equation are always dynamical:
\be
\dot{u} = - 3H_u, \hs{2} H_u^2 = H^2 - \frac{V}{3} > 0, \hs{1} \mbox{for all $V < 0$}.
\label{s7a.10}
\ee
In particular
\be
h_1^2 = h_0^2 + \frac{|\ve|}{3} > 0,
\label{s7a.11}
\ee
hence it is guaranteed that  $h_1 \neq 0$. The solution for $H$ and for $\dot{u}$ is formally the same 
as in eqs.\ (\ref{s7a.9}), except that on has to replace $\ve = -|\ve|$. The restriction imposed by the 
single-valuedness of $H[u]$ now becomes
\be
\lh \frac{m u}{3} \rh^2 -  H^2 = \frac{|\ve|}{3},
\label{s7a.12}
\ee
which is a hyperbola with branches in the left and right quadrants, leaving an opening on the $u$-axis in the
interval 
\[
- \frac{\sqrt{|\ve|}}{m} < u < \frac{\sqrt{|\ve|}}{m}.
\]
This hyperbola is shown in fig.\ 3; passing through the allowed interval on the $u$-axis, $H$ can cross from 
positive to negative values. As $H(t)$ is a non-increasing function, this will eventually happen and an expanding 
universe will turn into a contracting universe. We have seen this behaviour already in the example given by 
eqs.\ (\ref{s4a.4}), (\ref{s4a.5}). We return to this point for a fuller discussion in the next section. 

\section{Series expansions: turning points \label{s8}}

As eq.\ (\ref{1.8}) shows, in single scalar cosmology $\dot{H} \leq 0$. Now the scalar field $u = \sqrt{3/2}\, \vf$
can have turning points where $\dot{u} = H_u = 0$, but $\ddot{u} = - V_u \neq 0$. Adapting eq.\ (\ref{2.5}) to 
the present notation
\be
V_u = 6 H_u \lh H - H_{uu} \rh, 
\label{s8a.1}
\ee
hence such a turning point occurs if $H_{uu}$ is singular in such a way that at this point
\be
0 < \left| H_u H_{uu} \right| < \infty.
\label{s8a.2}
\ee
As discussed in the previous sections, in the $(u, H)$-plane turning points lie on the curves 
\[
V(u) = 3 H^2(u),
\]
which bound the domain of physically allowed values. In fact all points on these boundary curves 
are turning points, unless $V_u = 0$, i.e.\ at a local extremum of the potential. In the latter case a 
solution with $\dot{u} = H_u = 0$ can exist only if this extremum occurs at a non-negative value 
of $V$. 

In the neighborhood of a turning point $u(t)$ takes identical values before and after the turning point;
but as $\dot{u} = - 3 H_u \neq 0$ away from the turning point, and therefore $\dot{H} < 0$ both before 
and after, $H[u(t)]$ necessarily becomes double-valued there. As a result, the power series expansion 
studied in sect.\ \ref{s6} can not be used in this neighborhood.  

This double-valuedness can be resolved by a reparametrization of the field. For definiteness it is 
convenient to consider a point where $u$ reaches a maximum $u_m$; then a new dynamical variable 
$\eta(t)$ can be introduced such that in a sufficiently small but finite time interval around the turning point 
\be
u(t) = u_m - \eta^2(t).
\label{s8a.3}
\ee
At the turning point $\eta = 0$, and the evolution of $u(t)$ can be parametrized by a monotonically increasing 
function $\eta(t)$,  the negative and positive values of $\eta$ corresponding to $u$ before and after the 
turning point, respectively. In case $u_m$ were a minimum, we could similarly define 
\be
u(t) = u_m + \eta^2(t).
\label{s8a.4}
\ee
However, for our discussion we will assume a maximum and use eq.\ (\ref{s8a.3}). 

We first re-express our dynamical equations in terms of the new field $\eta(t)$. Using the definition
(\ref{s8a.3}) and the notation $H_{\eta} = dH/d\eta$, it is straightforward to derive the equations
\be
H^2 - \frac{1}{4\eta^2}\, H_{\eta}^2 = \frac{1}{3}\, V[u(\eta)], \hs{2}
\dot{\eta} = - \frac{3}{4\eta^2}\, H_{\eta}.
\label{s8a.5}
\ee
These equations can be used to develop a new power series expansion
\be
H = \sum_{n \geq 0} g_n \eta^n, \hs{2} H_{\eta} = \sum_{n\geq 0} (n+1) g_{n+1} \eta^n.
\label{s8a.6}
\ee
Now expressed in $\eta$ the condition (\ref{s8a.2}) at the turning point, where $H_u = 0$, reads
\be
\ld H_{\eta} \right|_{\eta = 0} = 0, \hs{2} 0 < \left| \frac{1}{\eta^3} H_{\eta} H_{\eta\eta} \right|_{\eta = 0} < \infty.
\label{s8a.7}
\ee
In terms of the expansions (\ref{s8a.6}) this is translated as 
\be
g_1 = g_2 = 0.
\label{s8a.8}
\ee
As a result
\be
H = g_0 + g_3 \eta^3 + g_4 \eta^4 + 5 g_5 \eta^5 ..., \hs{2} 
H_{\eta} = 3 g_3 \eta^2 + 4 g_4 \eta^3 + 5g_5 \eta^4 +  ...
\label{s8a.9}
\ee
Also, assuming that the potential $V$ has a power series expansion (\ref{s6a.8}) with $u_0 = u_m$,
the potential has the expansion
\be 
V = \sum_{n \geq 0} (-1)^n v_n\, \eta^{2n}. 
\label{s8a.10}
\ee
Then eqs.\ (\ref{s8a.5}) imply for the coefficients $g_n$:
\be
\ba{l}
\dsp{ g_0^2 = \frac{v_0}{3}, \hs{2} g_1 = g_2 = 0, \hs{2} g_3^2 = \frac{4v_1}{27}, }\\ 
 \\
\dsp{ g_4 = \frac{g_0}{3}, \hs{2} g_5 = \frac{g_3}{30 v_1} \lh 2 v_0 - 9 v_2 \rh, }\\
 \\
\dsp{ g_6 = \frac{2g_0}{405 v_1} \lh - 2 v_0 + 9 v_2 \rh, \hs{1} g_7 = \frac{2g_3}{21} 
 \left[ 1 + \frac{9v_3}{4v_1} + \frac{1}{240 v_1^2} \lh 2 v_0 - 9 v_2 \rh \lh 2v_0 + 15 v_2 \rh \right], }
\ea
\label{s8a.11}
\ee
etc. Observe, that these equations require $v_0 \geq 0$ and $v_1 \geq 0$. The first condition implies
that turning points only occur at non-negative values of the potential $V$. The second condition is a 
direct consequence of our choice to consider a turning point which is a maximum of $u$: 
\[
V_u(u_m) = v_1 \geq 0.
\]
The equations for $H[\eta]$ and $\eta(t)$ now become
\be
\ba{lll}
H & = & \dsp{ \sqrt{\frac{v_0}{3}} \left[ 1 - \frac{2}{3} \sqrt{\frac{v_1}{v_0}}\, \eta^3 + \frac{1}{3}\, \eta^4 
 + \frac{9 v_2 - 2 v_0}{45 \sqrt{v_0 v_1}}\, \eta^5 + ... \right], }\\
 & & \\
\dsp{ \dot{\eta} }& = & \dsp{ \sqrt{\frac{3v_1}{4}} \left[  
 1 + \frac{2}{3} \sqrt{\frac{v_0}{v_1}}\, \eta + \lh \frac{2v_0 - 9 v_2}{18 v_1} \rh \eta^2 + ... \right] }
\ea
\label{s8a.12}
\ee
Note, that in order to get a monotonically increasing $\eta(t)$, we have to take the negative 
square root for $g_3$:
\[
g_3 = - \frac{2}{3} \sqrt{\frac{v_1}{3}}.
\] 

In the present formulation one can derive yet another formula for the total expansion factor in  a given 
time interval $(t_1, t_2)$:
\be
\ba{lll}
N & = & \dsp{ \int_1^2 H dt = - \frac{4}{3}\, \int_1^2 d\eta\, \frac{\eta^2 H}{H_{\eta}} }\\
 & & \\
 & = & \dsp{- \frac{4}{3}\,  \int_1^2 d\eta\, \frac{g_0 + g_3 \eta^3 + g_4 \eta^4 + g_5 \eta^5 + ...}{
  3g_3 + 4 g_4 \eta + 5 g_5 \eta^2 + ...}. }
\ea
\label{s8a.13}
\ee
The result can be written in a series expansion as
\be
N = \left[ \nu_1\, \eta + \nu_2\, \eta^2 + \nu_3\, \eta^3 + \nu_4\, \eta^4 + .... \right]_1^2, 
\label{s8a.14}
\ee
with coefficients given by
\be
\ba{ll}
\dsp{ \nu_1 = - \frac{4}{9} \frac{g_0}{g_3}, }& \dsp{ \nu_2 = \frac{8}{27} \frac{ g_0 g_4}{g_3^2}, }\\
  \\
\dsp{ \nu_3 = \frac{20}{81} \frac{g_0g_5}{g_3^2} - \frac{64}{243} \frac{g_0g_4^2}{g_3^3}, }&
\dsp{ \nu_4 = - \frac{1}{9} + \frac{2}{9} \frac{g_0g_6}{g_3^2} - \frac{40}{81} \frac{g_0g_4g_5}{g_3^3}
 + \frac{64}{243} \frac{g_0g_4^3}{g_3^4}. }
\ea
\label{s8a.15}
\ee
Note again, that by taking the negative value for $g_3$ the first coefficient becomes positive:
\be
\nu_1 = \frac{2}{3} \sqrt{\frac{v_0}{v_1}}.
\label{s8a.16}
\ee
 
\section{Quadratic potentials \label{s9}}

The general description of turning points developed in sect.\ \ref{s8} can be illustrated with the example 
of quadratic potentials considered earlier
\[
V = \ve + \frac{m^2}{3}\, u^2.
\]
We consider a solution $u(t)$ which rolls down the potential from negative values to positive values,
reaching a turning point at some positive $u_m > 0$, where $\dot{u}_m = 0$ and $V_u(u_m) > 0$. 
Using the parametrization (\ref{s8a.3}) the potential is expressed as 
\be
V = v_0 - v_1 \eta^2 + v_2 \eta^4, \hs{2} v_0 = \ve + \frac{m^2}{3} u_m^2, \hs{1} 
v_1 = \frac{2m^2}{3}\, u_m, \hs{1} v_2 = \frac{m^2}{3}.
\label{s9a.1}
\ee
The requirement $v_1 > 0$ is satisfied automatically, whilst the condition $v_0 > 0$ is non-trivial only 
if $\ve < 0$, requiring $u_m$ to be in the domain of  $V(u_m) > 0$. As before we distinguish the cases
$\ve \geq 0$ and $\ve < 0$. 

For non-negative $\ve$ the coefficients $g_n$ in eq.\ (\ref{s8a.11}) become
\be
\ba{ll}
\dsp{ g_0 = \frac{mu_m}{3} \sqrt{1 + \frac{3\ve}{m^2 u_m^2}}, }&
\dsp{ g_3 = - \frac{2m}{9} \sqrt{2u_m}, }\\
 & \\
\dsp{ g_4 = \frac{mu_m}{9} \sqrt{1 + \frac{3\ve}{m^2 u_m^2}}, }& 
\dsp{ g_5 = \frac{m}{15 \sqrt{2u_m}} \lh 1 - \frac{2u_m^2}{9} - \frac{2\ve}{3m^2} \rh, }\\
 & \\
\dsp{ g_6 = \frac{m}{135} \sqrt{1 + \frac{3\ve}{m^2 u_m^2}} \lh 1 - \frac{2u_m^2}{9} - \frac{2\ve}{3m^2} \rh, }&
 ...
\ea
\label{s9a.2}
\ee
The singularity of $g_5$ for $u_m \rightarrow 0$ indicates, that the only consistent solution with $\dot{u} = 0$
at the point $u = 0$ is the one with constant $H = g_0 = \sqrt{\ve/3}$, corresponding to a de Sitter space for 
$\ve > 0$, or a Minkowski space for $\ve = 0$. This is a stationary solution, rather than a turning point.

For negative $\ve$ turning points can occur, and eqs.\ (\ref{s9a.2}) still hold, if $u^2_m > 3 |\ve|/m^2$. 
Therefore, if the scalar field starts at a large enough value in the region $V(u) > 0$, it can roll down the 
potential and oscillate, meeting a number of turning points, until it can no longer escape from the region 
$V < 0$  and the universe starts to contract back to infinitely small size. The point is, that after each turning 
point the Hubble parameter will continue to decrease: $\dot{H} < 0$, until it finally crosses over into the region 
of negative $H$ and the contraction phase starts. 

\section{Discussion and conclusions \label{s10}}

The main concern of this paper is the cosmological evolution of spatially flat universes driven by a single
scalar field. Such a scenario may be relevant for the present epoch in the evolution of our accelerating 
universe, and it may have a bearing on the vey early universe going through an epoch of inflation. 
Observations of the CMB require an inflationary expansion by at least 60-70 $e$-folds, which can happen
only if the universe spends a relatively long time in a phase with large Hubble parameter.  This 
requirement is usually expressed by the slow-roll condition; therefore it is of interest to study this 
condition in the present context of single scalar-field cosmology. 

The accelaration of the universe can be expressed as
\be
\frac{\ddot{a}}{a} = \dot{H} + H^2 = H^2 \lh 1 - \eps \rh,
\label{s10a.0}
\ee
where the slow-roll parameter is defined by 
 \be
 \eps =  \frac{3 H_u^2}{H^2}.
 \label{s10a.1}
 \ee
Thus an accelerated expansion requires $0 \leq \eps < 1$. Now we can combine the two 
inequalities 
\be
H^2 > 3 H_u^2, \hs{2} 3H^2 - V = 3H_u^2 \geq 0, 
\label{s10a.2}
\ee
to translate the slow-roll condition to a double bound on $H^2$:
\be
\frac{V}{3} \leq H^2 < \frac{V}{2}.
\label{s10a.3}
\ee
Clearly these bounds can be satisfied only in a region where $V > 0$. Also, the bound is 
always satisfied at and near a turning point where $H_u = 0$. In terms of our series expansion
(\ref{s8a.9}) this is at $\eta = 0$, where $H^2 = V/3$. In order to estimate the total expansion 
factor in the slow-roll domain (\ref{s10a.3}) near a turning point, we also ought to find a value 
for $\eta$ at $H^2 = V/2$ for a given specific potential. In general however, we can use the
observation that in Planck units $u_m$ and $\eta$ must satisfy 
\be
\eta^2 < u_m < 1,
\label{s10a.4}
\ee
and as a first approximation 
\[
N = - \frac{4}{9} \frac{g_0}{g_3} \int_0^{\eta_m} d\eta \lh 1 - \frac{4}{3} \frac{g_4}{g_3} \eta + ... \rh 
 \approx \frac{2}{3} \sqrt{\frac{v_0}{v_1}} \times {\cal O}(1),
\]
where $\eta_m$ is the upper limit for $\eta$ where $H = V/2$. The condition on the number of 
$e$-folds for inflation then becomes
\[
\frac{v_0}{v_1} \sim 10^4.
\]
However, significant modification of this estimate may result for specific potentials 
\ct{damour-mukhanov}. 

The methods we have used in this paper to find solutions for the the cosmological equations of a 
spatially flat universe driven by scalar fields relies heavily on the fact that we have assumed a single 
field to drive the cosmological expansion. This allows us to replace time by the field as the evolution 
parameter. Actually eq.\  (\ref{1.10}) suggests another option, taking $X^0 = \sqrt{6} \ln a$ as evolution 
parameter; this might be more readily generalizable to the case of many scalar fields 
\ct{grootnib-vantent2002, vantent2002}. At present such a modification of the methods presented 
here is under investigation. 
\vs{3}

\nit
{\bf Acknowledgement} \\
Discussions with Reinier de Adelhart Toorop in an early stage of this work are gratefully acknowledged. 
This work is part of the research program of the Foundation for Fundamental Research of Matter (FOM).

\np

\end{document}